\documentclass[%
aip,
rsi,%
amsmath,amssymb,
reprint,%
]{revtex4-1}
\usepackage{graphicx}
\usepackage{dcolumn}
\usepackage{bm}
\usepackage{amsmath}
\usepackage{color,soul}
\usepackage{natbib}
\usepackage{gensymb}
\everymath{\displaystyle}

\begin{document}


\title{Measurement of anisotropic thermal conductivity of a dense forest of nanowires using the $3\omega$ method} 
\vspace{0.5cm}


\author{Dhruv Singhal}
\email{dhruv.singhal@cea.fr}
\affiliation{Univ. Grenoble Alpes, CEA, INAC-Pheliqs, 38000 Grenoble, France}
\affiliation{Institut N\'{e}el, CNRS, 38000 Grenoble, France}
\affiliation{Univ. Grenoble Alpes, Grenoble, France}
\author{Jessy Paterson}%
\affiliation{Institut N\'{e}el, CNRS, 38000 Grenoble, France
}%
\affiliation{Univ. Grenoble Alpes, Grenoble, France
}%
\author{Dimitri Tainoff}%
\affiliation{Institut N\'{e}el, CNRS, 38000 Grenoble, France
}%
\affiliation{Univ. Grenoble Alpes, Grenoble, France
}%
\author{Jacques Richard}
\affiliation{Institut N\'{e}el, CNRS, 38000 Grenoble, France}

\author{Meriam Ben-Khedim}

\affiliation{Institut N\'{e}el, CNRS, 38000 Grenoble, France}

\affiliation{Technology R\&D, STMicroelectronics, 13106 Rousset, France}

\author{Pascal Gentile}
\affiliation{Univ. Grenoble Alpes, CEA, INAC-Pheliqs, 38000 Grenoble, France}

\author{Laurent Cagnon}%
\affiliation{Institut N\'{e}el, CNRS, 38000 Grenoble, France
}%
\affiliation{Univ. Grenoble Alpes, Grenoble, France
}%
\author{Daniel Bourgault}%
\affiliation{Institut N\'{e}el, CNRS, 38000 Grenoble, France
}%
\affiliation{Univ. Grenoble Alpes, Grenoble, France
}%

\author{Denis Buttard}%
\affiliation{Univ. Grenoble Alpes, CEA, INAC-Pheliqs, 38000 Grenoble, France
}%
\affiliation{Univ. Grenoble Alpes, Grenoble, France
}%
\author{Olivier Bourgeois}%
\email{olivier.bourgeois@neel.cnrs.fr}
\affiliation{Institut N\'{e}el, CNRS, 38000 Grenoble, France
}%
\affiliation{Univ. Grenoble Alpes, Grenoble, France
}%


\begin{abstract}
The $3\omega$ method is a dynamic measurement technique developed for determining the thermal conductivity of thin films or semi-infinite bulk materials. A simplified model is often applied to deduce the thermal conductivity from the slope of the real part of the ac temperature amplitude as a function of the logarithm of frequency, which in-turn brings a limitation on the kind of samples under observation. In this work, we have measured the thermal conductivity of a forest of nanowires embedded in nanoporous alumina membranes using the 3$\omega$ method. An analytical solution of 2D heat conduction is then used to model the multilayer system, considering the anisotropic thermal properties of the different layers, substrate thermal conductivity and their thicknesses. Data treatment is performed by fitting the experimental results with the 2D model on two different sets of nanowires (silicon and BiSbTe) embedded in the matrix of nanoporous alumina template, having thermal conductivities that differ by at least one order of magnitude. These experimental results show that this method extends the applicability of the $3\omega$ technique to more complex systems having anisotropic thermal properties.
\end{abstract}


\maketitle 
\section{Introduction}
Thermal management is a key issue in microelectronic industry for several essential cases specially heat dissipation
 (heat pipes, heat sink), local cooling and energy recovery either to prevent premature aging or to do self powering of sensors\cite{Liu2014,richter2017thermal,6962715}. With the decrease in the size of the electronic components, the number of experimental developments dedicated to thermal characterization of nanomaterials has significantly increased over the last decade\cite{cahill2014nanoscale,volz2016nanophononics}. Using nano-structuration to reduce the phonon contribution to the thermal conductivity is seen as the best way for improving the efficiency of thermoelectrics \cite{pernot2010precise,duchemin2011atomistic,heron2010blocking}. Moreover, increased interest in the research field of energy harvesting involving the innovative concepts of thermal management applied to thermoelectric generators has been seen\cite{perez2014,aswal2016key}. With complex three-dimensional nano-architectures used for energy harvesting, precise measurement of thermal conductivity is a crucial issue for appropriate fabrication of nano-devices and it still remains a challenge at small length scales\cite{hippalgaonkar2010fabrication,Gnam,CRAC}.

Among nanostructured materials, nanowires (NWs) have shown a great potential in thermoelectric applications with remarkable increase in the figure of merit\cite{chen2003recent}. Matrix of highly-dense nanowires opens up the possibility of chip-level thermoelectric generators which can be integrated in existing electronic circuitry with high scalability\cite{li2011chip}. Structures made with nanowires embedded in templates (which are used for the growth of nanowires through bottom-up fabrication technique) have more mechanical strength and offer the possibility of having nanowires with high aspect-ratios. The presence of the template will affect the measurement of the thermal properties of the film of nanowires due to parasitic thermal paths. The challenge is the measurement of thermal conductivity of anisotropic films with high difference between the in-plane and cross-plane thermal conductivities.
All these factors bring in the need for precise measurements of the thermal properties of these complex anisotropic films of nanowires embedded in templates for their potential use in modern thermoelectric device.  

Several techniques have been developed in the past for the measurement of the in-plane thermal conductivity\cite  {abrosimov1969thermal,nath1973experimental,boikov1975methods,kelemen1976pulse,boiko1973method,zink2004specific,sultan2009thermal,bourgeois2007measurement,sikora2012highly} and cross-plane thermal conductivity\cite{cahill1987thermal,cahill1989thermal,duquesne2009thermal} that use the principles of the 3$\omega$ method\cite{dames20051,liu2014sensitive,lee1997thermal}.  
Regarding the measurement of single nanowire, various techniques have been used for measuring the thermal conductivity like the double-platform or the Scanning Thermal Microscopy (SThM) \cite{hippalgaonkar2010fabrication,hochbaum2008enhanced,munoz2013fab}. However, the experimental demonstration of a 3$\omega$ method allowing the measurement of a macroscopic array of nanowires is still awaited. 

Among all the techniques at disposal, the 3$\omega$ is a well-developed method for measurement of thermal conductivity of thin films, nanosystems and bulk materials\cite{cahill1990thermal,bourgeois2007measurement,sikora2012highly}. It has significant advantages by facilitating the measurement of macroscopic systems, eliminating the errors arising from the black-body radiation and reducing noise with the help of the experimental set-up dedicated for dynamical signal. It has been reported for acquiring the strong thermal anisotropy in nanoporous silicon\cite{kim2015strong} and 2D materials\cite{jang2015anisotropic,lee2015anisotropic,borca2005anisotropic}. This technique seems to be perfectly adapted for the measurement of thermal conductivity of a highly-dense array of nanowires embedded in a alumina template; especially in the case of thin film with anisotropic thermal transport properties. 

Here, we report the thermal conductivity measurements on large assemblies of nanowires embedded in nanoporous alumina templates using the 3$\omega$ method. These large assemblies of nanowires lead to anisotropic 2D heterostructures, whose derivation of thermal conductivity through experimental ways is not straightforward. A detailed description of the fabrication of two different types of nanowires, namely Bi$_{0.37}$Sb$_{1.49}$Te$_{3.14}$ and silicon, is presented. Since the sample structures are complex, the simplified analytical solution for the 3$\omega$ method can no longer be applied to deduce the thermal conductivity of the nanowires. Moreover, the embedded-nanowire film has anisotropic thermal properties, meaning, the in-plane and the cross-plane thermal conductivities differ. We show experimentally that the two-dimensional heat conduction model proposed by Borca-Tasciuc~\textit{et al.} in 2001 can be used for the general case of a multilayer film-on-substrate system with anisotropic thermophysical properties for the extraction of thermal conductivity of the nanowires\cite{borca2001data}. The measurements are done on approximately $5\times 10^{5}$~NWs simultaneously\footnote{The number of nanowires has been estimated from the SEM image of the sample as the number of nanowires located under the transducer.}.

Measurement on such large number of nanowires reduces significantly the error arising from the existence of thermal contact resistances during the single nanowire measurements or from possible dispersion of properties from one nanowire to another. We finally demonstrate that using this method the thermal properties of nanowires fabricated in a bottom-up technique can be obtained with reasonable sensitivity.

\section{Experimental Details}
Highly dense array of nanowires ($10^9$/cm$^2$) are fabricated with a bottom-up growth technique. In order to have a large density of nanowires with high aspect ratio, nanoporous alumina templates have been used for the growth of nanowires. In order to perform the 3$\omega$ measurement, a thin layer (50~nm) of Al$_{2}$O$_{3}$ is deposited on the top of the sample for electrical isolation of the transducer from the mattress of nanowires that are usually electrically conductive. 
\begin{figure}[h]
	\includegraphics[width=0.3\textwidth]{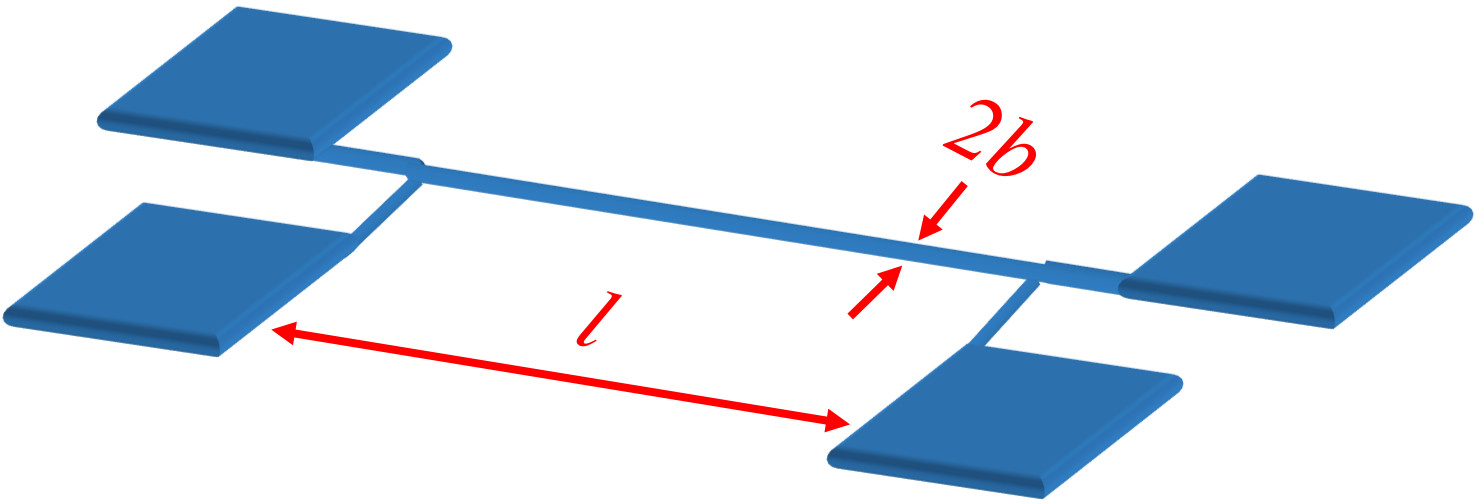}
	\caption{Platinum transducer in four probe geometry used for the 3$\omega$ measurements. The width of the transducer is defined by $2b$ (30 $\mu$m for BiSbTe and 5$\mu$m for silicon) and the length is referred to by $l$ (2 mm). The four pads are of the dimensions of $\mathrm{400 \mu m \times 400\mu m}$. }
	\label{fourprobe}
\end{figure}

With the help of laser lithography and magnetron sputtering, a platinum transducer (120~nm thick) is deposited on the top surface with a four probe geometry (Fig.~\ref{fourprobe}). Thermal conductivity measurements are done on the array of nanowires embedded in nanoporous Anodic Alumina Oxide (AAO) template. Two kinds of nanowire forests have been chosen for their very different bulk thermal transport properties to validate our approach: silicon nanowires as a highly conductive material and BiSbTe nanowires as a very low conductive materials. Indeed, the difference in the magnitude of the bulk thermal conductivities between both the materials is of two orders of magnitude ($\mathrm{\kappa_{Si} = 148~W m^{-1} K^{-1}, \kappa_{Bi_{0.5}Sb_{1.5}Te_{3}\bot}  = 1.75 ~W m^{-1} K^{-1}}$ $\mathrm{\kappa_{Bi_{0.5}Sb_{1.5}Te_{3}||}}  \mathrm{ ~ = 0.75~ W m^{-1} K^{-1}}$; $\bot, ||$ are the crystallographic directions of the material.)\cite{caillat1992thermoelectric,poudel2008high,grasso2013ultra}. This will allow to probe the technique over a broad range of thermal conductivity in two very different conducting limits.

\subsection{Sample Fabrication}
AAO was fabricated by anodizing aluminum by a two-step anodization technique \cite{masuda1996fabrication} in 3\% weigh/volume (w/v) oxalic acid solution. The diameter and the interpore distance is tuned with the help of anodizing potential and the concentration of the acid.  

\subsubsection{Fabrication of BiSbTe NWs}
Aluminum foil, 99.99\% pure, is used for the fabrication of nanoporous alumina template with the applied anodizing voltage of 40~V in 3\% w/v oxalic acid solution. Pore widening is done by dipping the AAO template in 5\% phosphoric acid solution at 35$\degree$C for six minutes. The resulting pore diameter is 60$\pm$5~nm.  The p-type Bi$_{0.37}$Sb$_{1.49}$Te$_{3.14}$ NW array is fabricated by pulse electro-deposited in the AAO at room temperature with the applied potential of -0.3~V\cite{Meriam}.   

\begin{figure}[h]
	\includegraphics[width=0.4\textwidth]{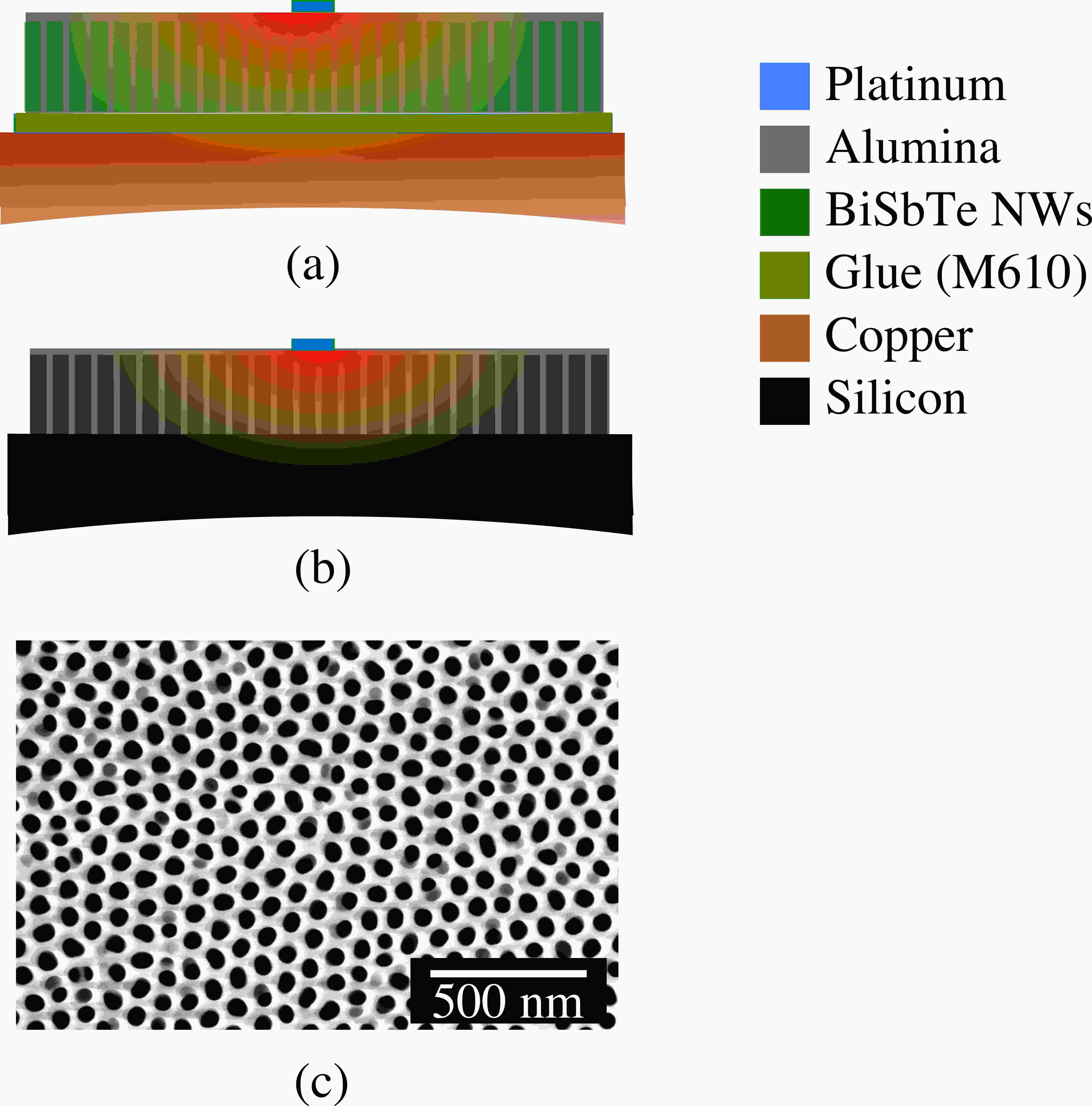}
	\caption{Schematic diagrams of the samples showing the multilayer system composed of the nanowire mattress, the insulating layer (alumina) on top and the platinum thermometer. (a) Bi$_{0.37}$Sb$_{1.49}$Te$_{3.14}$ nanowires embedded in AAO and, (b) Si NWs embedded in AAO. (c) is a Scanning Electron Microscope (SEM) picture from the top of the AAO template used for the growth of the nanowires.}
	\label{sketch}
\end{figure}

\subsubsection{Fabrication of silicon NWs}
Aluminum (10$\pm$0.05$\mu$m) is deposited on a highly p-doped silicon wafer $<100>$ (750~$\mu$m thick) by Electron Beam Physical Vapor Deposition (EBPVD). Anodization is carried out by two-step anodisation technique\cite{masuda1996fabrication} at room temperature with 3\% w/v oxalic acid solution. Applied potential of 40~V leads to the pore diameter of 65$\pm$5~nm. A pore widening step with 5\% phosphoric acid at 35~$\degree$C was performed to open the pore uniformly. At the end of the anodization process, the silicon wafer is oxidized at the interface of the aluminum and silicon. The SiO$_{2}$ present at the bottom of the pores is etched by exposing the sample to HF vapor. The post-treatment of AAO is followed by pulsed electro-deposition of gold at bottom of the pores which would act as a catalyst for the Vapor-Liquid-Solid growth mechanism in Chemical Vapor Deposition.  The amount of gold deposition at the bottom of the pores depends on the diameter of the pores.  The SiNWs are grown in a LPCVD (EasyTube~300 First Nano, a Division of CVD Equipment Corporation\textsuperscript{\textregistered}) operating at 580$\degree$~C and 6 torr of total pressure (SiH$_4$/HCl/N$_2$ - 50/100/1200 sccm respectively). The growth rate of the nanowires inside the porous alumina templates is 100-110 nm min$^{-1}$.

\subsection{Description of the 3$\omega$ measurement method}
The 3$\omega$ method is very well known technique for the measurement of thermal conductivity of thin films and bulk materials\cite{cahill1987thermal,cahill1989thermal,cahill1990thermal}. In this method, a metallic strip is deposited on top of the sample which acts as both the heater and the thermometer (commonly known as the transducer). The geometry of the full stack of layers is shown on Fig.~\ref{sketch}. The line transducer is biased by an ac current having an angular frequency $\omega$. This causes a temperature fluctuation due to Joule heating with an ac component at 2$\omega$ related to the thermal properties of the transducer and all its surroundings. The transducer resistance is then modulated at 2$\omega$ with an amplitude that depends on the temperature coefficient of resistance (TCR) of the transducer material: 
\begin{equation}
\alpha = \frac{1}{R}\frac{\partial R}{\partial T}
\label{alpha}
\end{equation}
The resulting 3$\omega$ voltage component contains the information about the thermal conductivity of the underlying layers. This method can be employed to measure the thermal conductivities of thin films down to several nanometers\cite{cahill1987thermal,cahill1989thermal,cahill1990thermal}. The accuracy in the determination of the thermal conductivity relies greatly on the heat conduction model applied. In general, an approximate analytical expression is employed based on the slope of the real part of the 3$\omega$ voltage expressed as a function of the logarithm of the electrical frequency to determine the thermal conductivity $k_{0}$:

\begin{equation}
\begin{split}
k_{0} &=-\frac{\alpha R^{2} I^{3}}{4 \pi l}\Bigg(\frac{\partial V_{3\omega - real}}{\partial \ln \omega}   \Bigg) ^{-1} \\
&=- \frac{1}{2\pi l} \Bigg[ \frac{\partial \Big(\displaystyle\frac{\Delta T_{real}}{P}\Big)}{\partial \ln \omega }\Bigg]^{-1}
\label{eq:TC}
\end{split}
\end{equation}

where, $\alpha$ is the temperature coefficient of resistance for the thermometer (obtained from Eq.~\ref{alpha}), $R$ is the resistance of the thermometer, $I$ is the applied rms current, $l$ is the length of the thermometer line, $V_{3\omega - real}$ is the in-phase component of the 3$\omega$ voltage, $\omega$ is the electrical angular frequency, $\Delta T_{real}$ is the in-phase temperature change and $P$ is the power dissipated by Joule heating.

\begin{figure}[h]
	\includegraphics[width=0.45\textwidth]{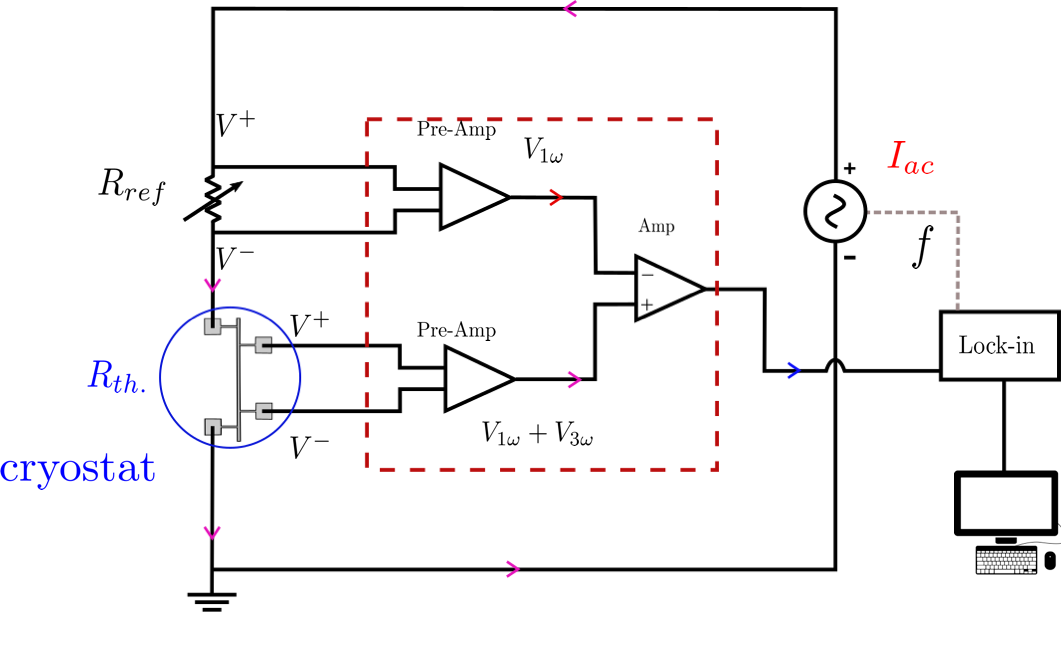}
	\caption{Schematic diagram of the electronic circuit used for the 3$\omega$ measurements. Signal is extracted from a four-probe transducer (highlighted in blue circle). The sample is kept in the cryostat under high vacuum ($10^{-5}$ mbar). A home made differential amplifier setup (highlighted in red rectangle) is used to extract the $V_{3\omega}$ out of the full voltage. The lock-in amplifier used in the experiments is a signal recovery model 7230 DSP by Ametek\textsuperscript{\textregistered}. The variable resistance, $R_{ref}$, is a programmable resistance model by IET\textsuperscript{\textregistered}, which has a precision of 0.001 $\Omega$. $I_{ac}$ is the ac current applied to the circuit with the frequency controlled by the lock-in amplifier.}
	\label{3omega}
\end{figure}

The electronic setup used in the 3$\omega$ measurements is schematically detailed in Fig.~\ref{3omega}. The samples are mounted in a cryostat and the temperature is controlled by the temperature controller TRMC2 developed at Institut N\'{e}el. Radiation shield and high vacuum ($<10^{-5}$ mbar) are used to minimize the losses by convection and radiation.

An alternating current $I_{ac}$ is applied to the platinum transducer within a large frequency range: between approximately 0.1~Hz and 1~kHz. The voltage from the transducer acquired by the lock-in amplifier contains both the first and third harmonics, however the $V_{3\omega}$ signal is significantly low in terms of amplitude as compared to the $V_{1\omega}$. Thus, the $V_{3\omega}$ is measured with the help of an active electronic setup used to increase the ${V_{3\omega}}/{V_{1\omega}}$ ratio. The two commonly used methods for the extraction of $V_{3\omega}$ signal are the differential bridge\cite{cahill1990thermal} (used in this work) and the Wheatstone bridge\cite{sikora2012highly}. In this experiments, the differential bridge has been preferred allowing a reduction of the $V_{1\omega}$ by a factor of ten to hundred\cite{liu2014sensitive}. Each data point is collected by averaging the lock-in amplifier reading over one second.

\subsection{2D thermal model}
\label{Model}
As mentioned before, most of the data treatment of 3$\omega$ experiment can be done by using the slope method, a one-dimensional heat conduction model to determine the thermal conductivity of thin films \cite{cahill1987thermal,cahill1989thermal,cahill1990thermal,kim1999application}. However, when the geometry of the sample becomes much more complex with multilayer on substrate and especially when layers with anisotropic thermophysical properties are present, a more general model of multilayer film-on-finite/semi-infinite substrate system  is required to determine the thermal conductivities\cite{lee1997thermal,chen1998heat}. Several models have been utilized in the past for extraction of thermal conductivity of anisotropic films on substrates and of substrate with finite thicknesses\cite{chen1998heat,kim1999application}. The most complete model given by Borca-Tasciuc \textit{et al.}\cite{borca2001data} derived the expression for the temperature rise in the general case of the multilayer film-on-finite/semi-infinite substrates based on a two dimensional heat conduction model across the system and a uniform flux boundary condition between the heater and the top film. Here, the data interpretation will be done using a 2D analytical method based on the classical separation of variables technique\cite{carslaw1959heat}. This gives equivalent results than the Integral Fourier Transformation Technique (IFTT) used by other group\cite{borca2001data}. We neglect the contributions from the thermal mass of the transducer and the thermal boundary resistance. The temperature rise (complex number) in the transducer which is dissipating a power $P/l$ (W/m) (the peak electrical power per unit length) is given by IFTT \cite{borca2001data}:

\begin{equation}
\Delta T = \frac{-P}{\pi l \kappa_{y_{1}}} \int_{0}^{\infty} \frac{1}{A_{1} B_{1}} \frac{\sin^{2}(b\lambda)}{b^{2}\lambda^{2}} d\lambda
\end{equation}

where,

\begin{equation}
A_{i-1} =\displaystyle \frac{A_{i}  \displaystyle\frac{\kappa_{y_{i}}B_{i}}{\kappa_{y_{i-1}}B_{i-1}}- \tanh(\phi_{i-1})}{1- A_{i}\displaystyle \frac{\kappa_{y_{i}}B_{i}}{\kappa_{y_{i-1}}B_{i-1}} \tanh(\phi_{i-1})} \ ,\quad i= 2... n 
\end{equation}

\begin{equation}
B_{i} = \Bigg( \kappa_{xy_{i}} \lambda^{2} + \frac{j~ 2 \omega}{\gamma_{y_{i}}} \Bigg)^{\frac{1}{2}}
\end{equation}

\begin{equation}
\phi_{i} = B_{i} d_{i} \,\qquad \kappa_{xy} = \frac{\kappa_{x}}{\kappa_{y}}
\end{equation}

where, $n$ refers to the total number of layers including the substrate with subscript $i$ for the $i^{th}$ layer from the top, 
subscript $x$ and $y$ correspond respectively to the direction perpendicular to the transducer and to the film/substrate, $b$ is the heater half width, $\kappa$ is the thermal conductivity,
$\kappa_{xy}$ is the anisotropy in thermal conductivity, it is the ratio of in-plane to the cross-plane thermal conductivity, $d$ is the layer thickness and, $\gamma$ is the thermal diffusivity. The thermal anisotropy $\kappa_{xy}$ contains by definition the interface thermal resistance between the nanowire and the alumina matrix.
For the substrate layer, $i = n$ and for substrates with finite thickness, the value of $A_{n}$ depends on the boundary condition at the bottom of the substrate. For adiabatic boundary conditions $A_{n}=-\tanh(B_{n}d_{n})$ and for isothermal conditions, $A_{n}=-1/\tanh(B_{n}d_{n})$. $A_{n}=-1$ if the substrate is semi-infinite.

\section{Experimental results}
The 3$\omega$ experiments are carried out in the temperature range of 260-300~K and the fitting of the experimental data is done with the above-described model. The physical properties of all the layers play a crucial role in the fitting along with their precise thicknesses. In order to get the thermal properties of the embedded nanowires only, we need to get information about the thermal properties of the porous matrix that carries also part of the heat. The empty porous alumina templates were closed from the top with the help of thin layer (100 nm) of aluminum oxide deposited by Atomic Layer Deposition (ALD). Then, the thermal conductivity of empty nanoporous AAO membranes having the very same structure (size of pores, filling factors) has been measured using the same $3\omega$ method. 

\begin{figure}[h]
	\includegraphics[width=0.45\textwidth]{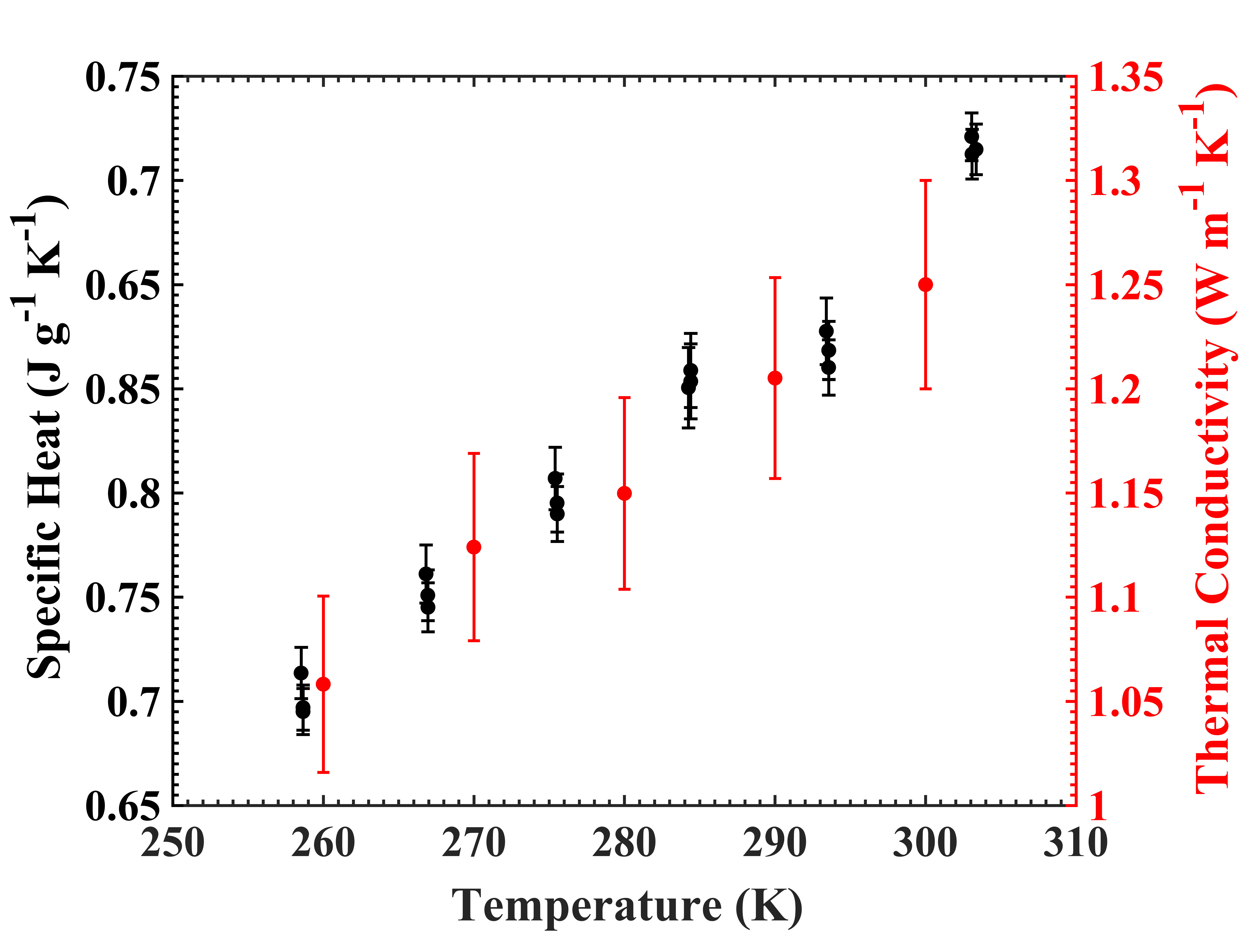}
	\caption{Variation of the specific heat and thermal conductivity with temperature of empty nanoporous alumina templates. The specific heat measurements are done with the help of the Physical Property Measurement System (PPMS) and the thermal conductivity is derived by fitting the measurements data obtained from the 3$\omega$ measurements with the model of empty nanoporous AAO. The error bars are estimated from the absolute error of the PPMS equipment used for the specific heat measurement and from the measurement of the 3$\omega$ voltage plus the error induced by the physical parameters used for the fitting of the final value of thermal conductivity.}
	\label{fig:alumina}
\end{figure}

In the Fig.~\ref{fig:alumina}, the variation with temperature of the specific heat of empty nanoporous AAO is plotted along with the thermal conductivity derived from fitting the above-mentioned model to the experimental data from the 3$\omega$ measurements. The cross-plane thermal conductivity was found to be 1.25~$\pm$~0.05~$\mathrm{W~m^{-1}~K^{-1}}$ at 300~K. The value is in accordance with the previous measurements done on similar structures (having a porosity of 30\% and pore diameter of 60~nm) at 300~K by Abad \textit{et al.}\cite{abad2016rules}. The value of thermal anisotropy in the modeling of nanoporous AAO was obtained using the formula given by Borca-Tasciuc \textit{et al.}\cite{borca2005anisotropic}: $\kappa_{xy} = \frac{\kappa_{x}}{\kappa_{y}} = \frac{1}{1 + \phi}$, where $\phi$ is the porosity. The error in the determination of thermal conductivity includes the absolute error arising from the 3$\omega$ measurements along with the error in parameters used in the data fitting. 
With these known thermal properties of the empty nanoporous AAO templates, measurements on the forest of nanowires of Bi$_{0.37}$Sb$_{1.49}$Te$_{3.14}$ and silicon can now be performed.

\subsection{Thermal conductivity of Bi$_{0.37}$Sb$_{1.49}$Te$_{3.14}$ nanowires}
\begin{figure}[h]
	\includegraphics[width=0.5\textwidth]{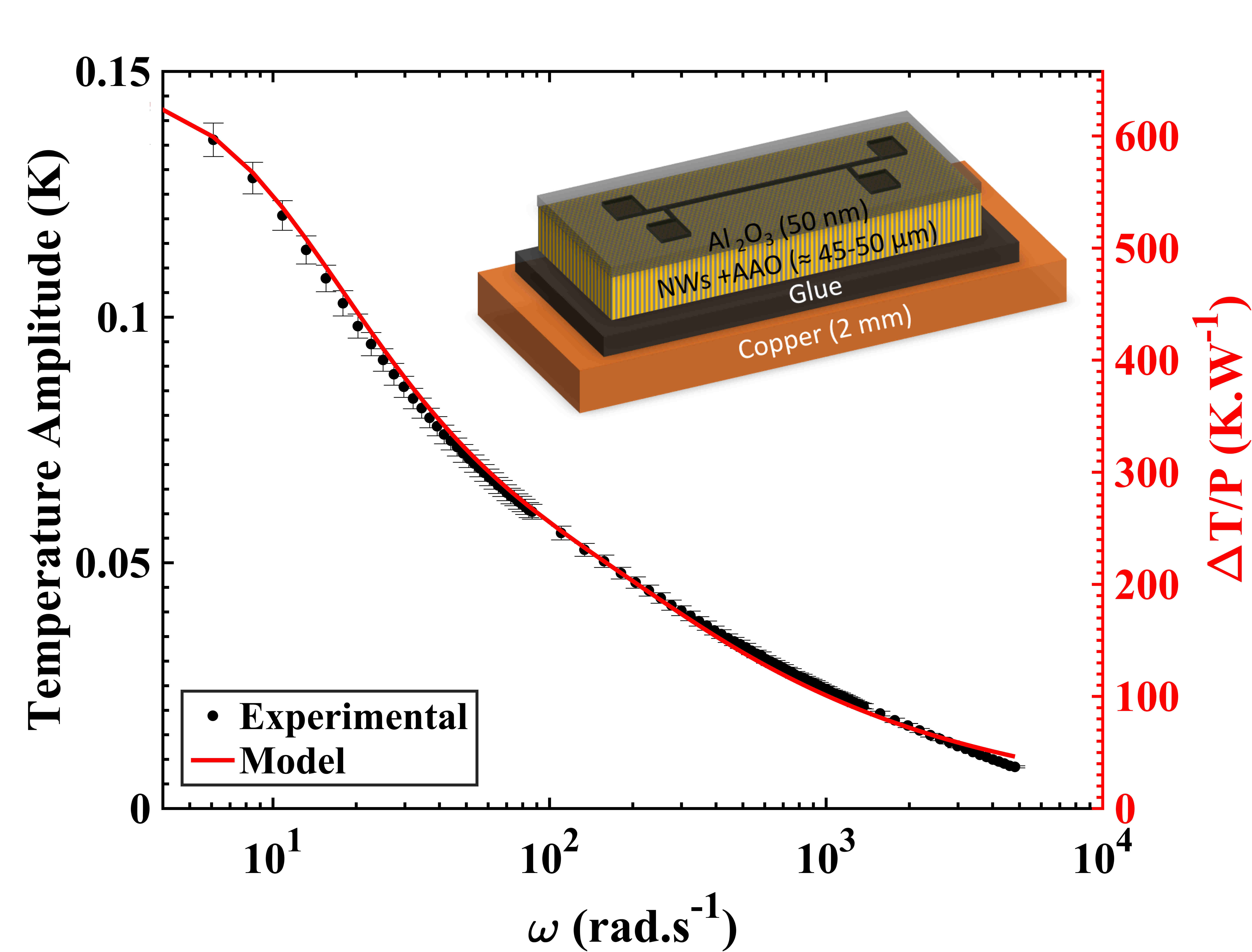}
	\caption{Amplitude of the temperature oscillation at 2$\omega$ with respect to the electrical angular frequency calculated from the 3$\omega$ voltage (black dots). The red line corresponds to the best fit of the experimental data based on the simulation done with the multilayer 2D model employed to extract the thermal conductivity of the forest of nanowires. The current used for the measurements is 2~mA and the width of the transducer is 30 $\mu$m. Inset: schematic of the sample with copper ($2$~mm) as substrate, glue to stick the membrane to substrate,  Bi$_{0.37}$Sb$_{1.49}$Te$_{3.14}$ NWs embedded in nanoporous alumina as the thin film ($\approx45~\mu$m), and a thin layer (50~nm) of $\mathrm{Al_2O_3}$. This layer serves as an electrical isolation between the thin film of NWs and AAO membrane and the platinum transducer on the top.}
	\label{BiSbTe}
\end{figure}

Thermal conductivity of highly dense forest of Bi$_{0.37}$Sb$_{1.49}$Te$_{3.14}$ nanowires has been measured with the 3$\omega$ method. In Fig.~\ref{BiSbTe} the modulus of temperature oscillation of the thermometer is shown as a function of the electrical angular frequency at 300~K. The changes in the slope are directly related to the thermal conductivity of the layers that are being probed at that particular frequency. Using the physical properties of all the multilayer system along with their appropriate thicknesses, simulation (red line) is performed based on the model presented above. The thermal conductivities are extracted from this operation as they are the only fitting parameters. 

 The cross-plane thermal conductivity of Bi$_{0.37}$Sb$_{1.49}$Te$_{3.14}$ nanowires extracted through the modeling is 0.57~$\pm$~0.02~$\mathrm{Wm^{-1}K^{-1}}$. In table~\ref{parameters}, all the physical parameters used in the data fitting are detailed. The thermal conductivity anisotropy $\kappa_{xy}$ in the layer of embedded nanowires is a fitting parameter. The thermal conductivities of the nanoporous alumina template and of the bismuth telluride nanowires are comparable, this is the reason why in-plane and cross-plane thermal transport are on the same order of magnitude ($\kappa_{xy}\sim 1$). 

\begin{table}[!]

	\begin{tabular}{|c|c|c|}
		\hline
		Sample Name (NWs)& Bi$_{0.37}$Sb$_{1.49}$Te$_{3.14}$ & Silicon\\
		\hline
		\hline
		 Porosity & 30$\pm$3\% & 30$\pm$3\% \\
		 \hline
		 Pore Filling & 95$\pm$2\% & 95$\pm$2\% \\
		 \hline
		 Anisotropy, $\kappa_{xy}$ & 0.95$\pm$0.03 & 0.48$\pm$0.04 \\
		\hline
		 Specific heat (J.g$^{-1}$.K$^{-1}$) & 0.29~$^($\cite{ftouni:tel-00995424}$^)$ &  0.86  \\
		\hline
		Density (g.m$^{-3}$)  & 6.73   &  2.32  \\
		\hline
	   AAO Thickness($\mathrm{\mu m}$)  & 45-50   &  8  \\
		\hline
		$\kappa$ ($\mathrm{W.m^{-1}.K^{-1}}$)  & 0.57$\pm$0.02   &  9.9$\pm$ 0.5 \\
		\hline
		
	\end{tabular}
	
	\caption{Physical parameters at 300~K used for fitting the experimental data using the model presented in the section~\ref{Model}. The estimation of porosity and filling factor have been obtained from the SEM analysis of the samples. The thermal conductivity of each sample at 300~K extracted from the 2D model is shown in the last row of the table.}
	\label{parameters}
\end{table}
 
\subsection{Thermal conductivity of silicon NWs}
\begin{figure}[h]
	\includegraphics[width=0.49\textwidth]{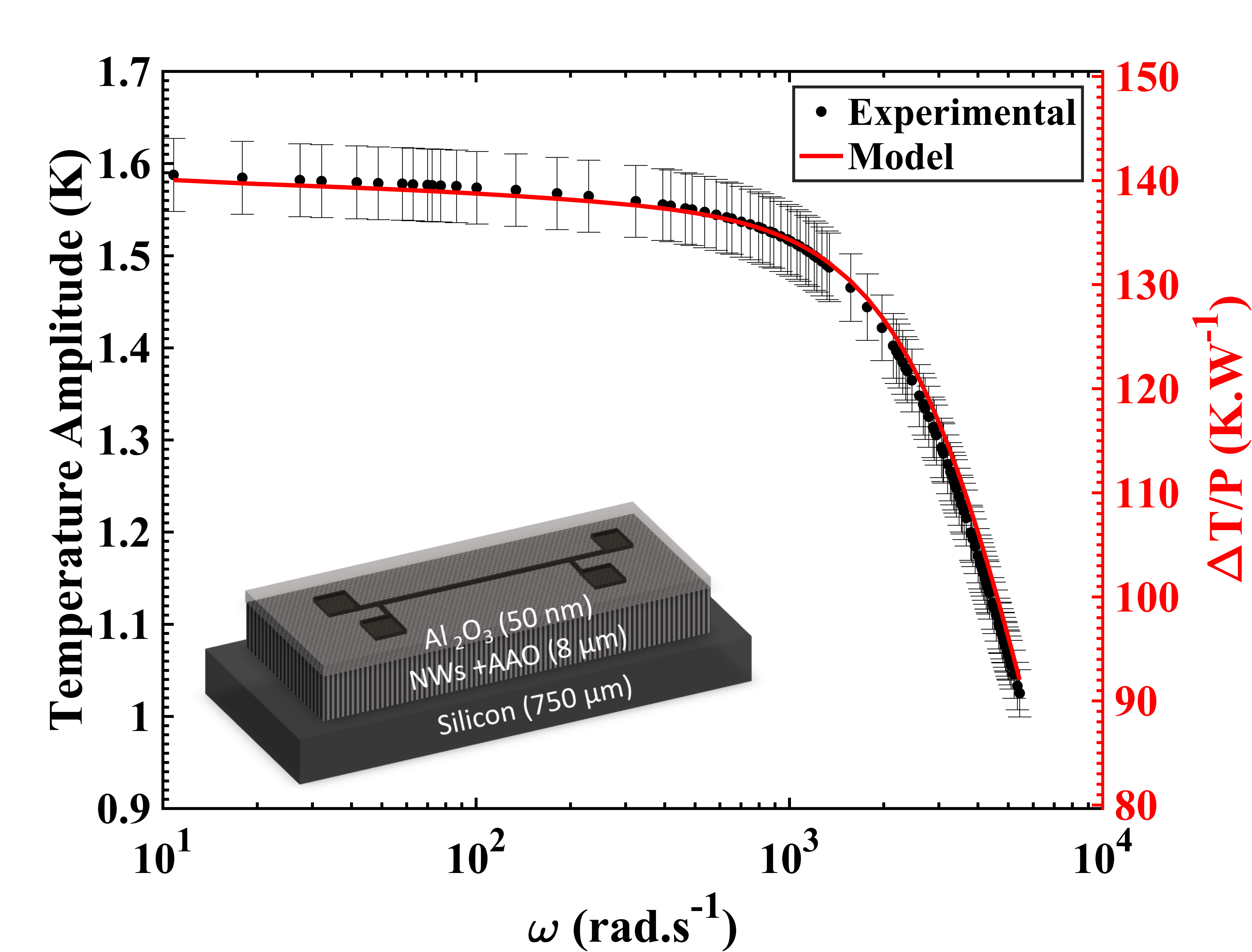}
	\caption{Amplitude of the temperature oscillation at 2$\omega$ with respect to the electrical angular frequency calculated from the 3$\omega$ voltage (black dots). Fitting of the experimental data based on the simulation done with the multilayer model employed to extract the thermal conductivity of the forest on nanowires (red line). The current used for the measurements is 4~mA and the width of the transducer is 5~$\mu$m. Inset: schematic of the sample with silicon ($750~\mu$m) as substrate, Si NWs embedded in nanoporous alumina as the thin film ($\approx 8~\mathrm{\mu}$m), and a thin (50~nm) of $\mathrm{Al_2O_3}$ to electrically isolate the platinum transducer (100~nm thick) on the top from the AAO thin film of NWs.}
	\label{SiNWs}
\end{figure}

The highly dense forest of Si nanowires has been also measured using the 3$\omega$ method. Fig.~\ref{SiNWs} shows the modulus of temperature oscillation of the thermometer with respect to the electrical angular frequency at 300~K. The difference in the frequency variation of the temperature oscillation from the one of the Bi$_{0.37}$Sb$_{1.49}$Te$_{3.14}$ NWs originates from the difference in the multilayer structure (Si NWs were grown directly on silicon wafers) but also from the high value of thermal conductivity of Si NWs. 

As the thermal conductivity of the nanoporous alumina is very small compared to silicon, very low thermal anisotropy is expected in the system, most of the heat conduction will be through the Si NWs which is the cross-plane direction of the film. Table~\ref{parameters} details all the physical parameters used for the fitting. The cross-plane thermal conductivity extracted through the modeling is 9.9~$\pm$~0.5~$\mathrm{Wm^{-1}K^{-1}}$ at 300~K\cite{PanAPL2015,hippal}. 

\begin{figure}[h]
	\includegraphics[width=0.5\textwidth]{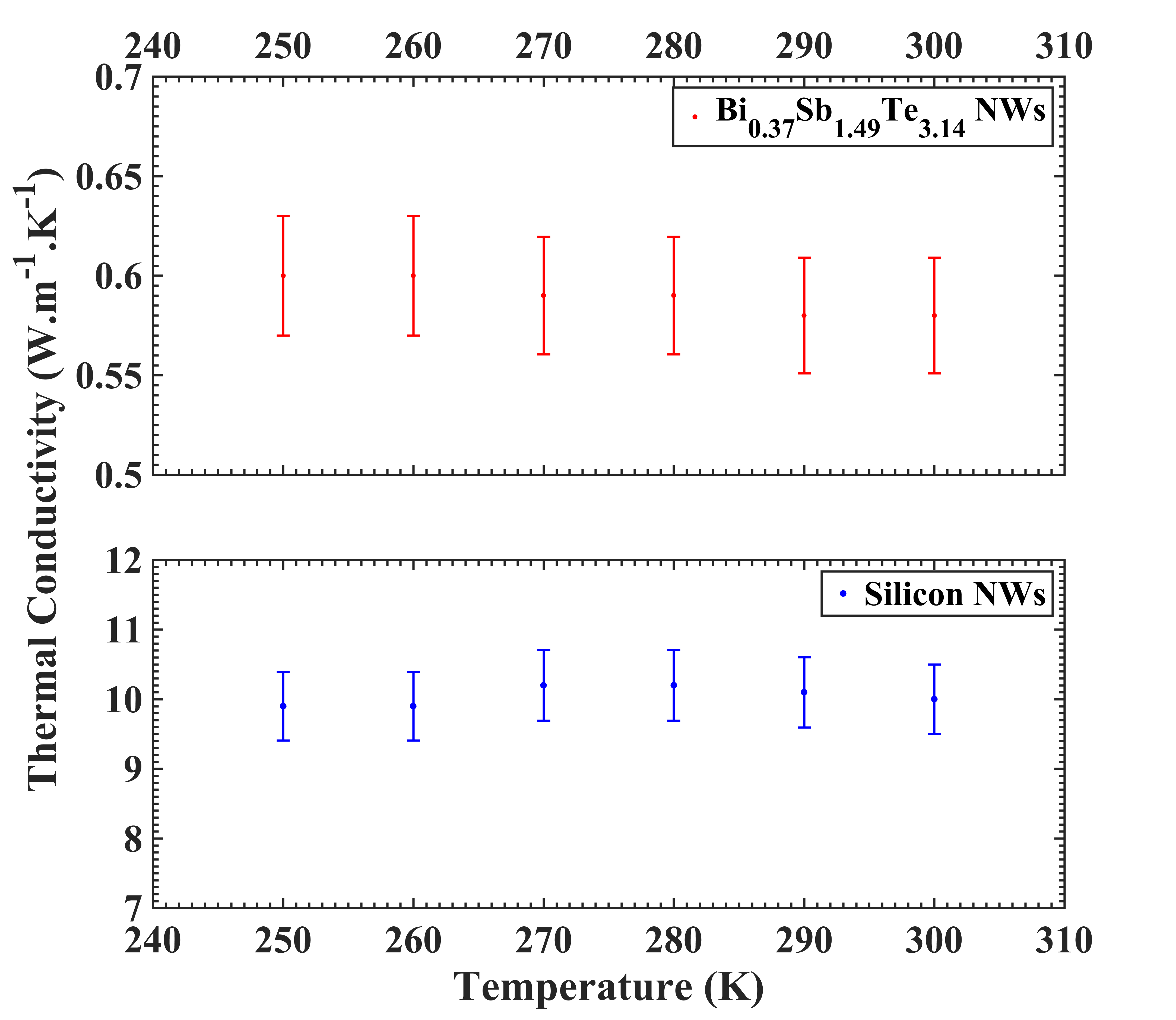}
	\caption{Temperature variation of thermal conductivities of Bi$_{0.37}$Sb$_{1.49}$Te$_{3.14}$ NWs and silicon NWs. It is extracted by fitting the experimental frequency variation of the 3$\omega$ voltage. The error bars are extracted from the best optimization of the fits.}
	\label{fig:TC}
\end{figure}

\subsection{Temperature variation of the thermal conductivity}

In Fig.~\ref{fig:TC}, the variation of the thermal conductivity is presented by fitting the experimental curves at different temperatures. The two materials have a difference of more than one order of magnitude in thermal conductivity at 300~K. 
This offers a validation of this 3$\omega$ techniques for determining thermal conductivity for anisotropic systems, especially for low conducting media. It shows that thermal conductivity of nanostructured materials can be extracted over a large temperature range from the proposed model. In terms of thermal transport, we can highlight the fact that in the case of bismuth telluride, the reduction of the size down to few tens of nanometers does not change significantly the thermal conductivity; this means that in bulk bismuth telluride, the phonon mean free path is already smaller than the size of the pores. This goes in opposition to silicon, for which, in the case of pores of 60~nm in diameter, the thermal conductivity has been divided by at least a factor of ten; the size reduction in diameter acting as an important source of phonon scattering\cite{ACSNano,CRAC}.

\section{Conclusion}
In this article, thermal conductivity measurements were done on systems that are thermally anisotropic: dense arrays of nanowires of Bi$_{0.37}$Sb$_{1.49}$Te$_{3.14}$ and silicon using the 3$\omega$ method. Nearly $5\times 10^{5}$~NWs were probed simultaneously during the measurements. The extraction of the thermal conductivity of the anisotropic films has been presented with the help of a two-dimensional heat conduction model. Thermal anisotropy along with the thermal conductivity of the films were determined with a high accuracy in the temperature range of 250~K to 300~K. The model has been applied to two materials with very different bulk thermal conductivities to prove the versatility of this data treatment method. This work demonstrates the possible extension of the applicability of the 3$\omega$ method to complex systems having thermal anisotropic layer or multilayers and consequently the measurement of a large number of nanowires.

\section*{Acknowledgements}
The authors would like to acknowledge the technical support and discussions provided by the Nanofab, P\^{o}le Capteur, P\^{o}le \'{E}lectronique and Cryogenic P\^{o}le facilities. We thank also the  people of Institut N\'{e}el, INAC and Plateforme Technologique Amont (PTA) in CEA. We acknowledge funding and support from the ANR project MESOPHON, the Laboratoire d'excellence LANEF in Grenoble (ANR-10-LABX-51-01), and from the ANRT. We would like also to thank ST Microelectronics for the collaboration on part of this work.

\bibliography{aipsamp}

\end{document}